\documentclass[12pt,preprint]{emulateapj}

\def\be{\begin{equation}}
\def\ee{\end{equation}}
\def\bea{\begin{eqnarray}}
\def\eea{\end{eqnarray}}

\usepackage{graphicx}

\begin{document}

\title{Short Gamma-ray Bursts: the mass of the accretion disk and the initial radius of the outflow}

\author{\sc Yi-Zhong Fan\altaffilmark{1,2} and Da-Ming Wei\altaffilmark{1,2}}
\altaffiltext{1}{Purple Mountain Observatory, Chinese Academy of Sciences, Nanjing 210008, China;}
\altaffiltext{2}{Key Laboratory of Dark Matter and Space Astronomy, Chinese Academy of Sciences, Nanjing 210008, China.}
\email{yzfan@pmo.ac.cn(YZF) and dmwei@pmo.ac.cn (DMW)}

\begin{abstract}
 In this work we estimate the accretion-disk mass in the specific scenario of binary-neutron-star-merger with  current observational data. Assuming that the outflows of short Gamma-ray Bursts (GRBs) are driven via neutrino-antineutrino annihilation we estimate the disk mass of about half of short bursts in the sample to be $\sim 0.01-0.1~M_\odot$, in agreement with that obtained in the numerical simulations. Massive disks ($\sim {\rm several}~ 0.1~M_\odot$) found in some other short GRBs may point to the more efficient magnetic process of energy extraction or the neutron star and black hole binary progenitor. Our results suggest that some short bursts may be really due to the coalescence of double neutron stars and are promising gravitational wave radiation sources. For future short GRBs with simultaneous gravitational-wave detections, the disk mass may be reliably inferred and  the validity of our approach will be tested. We also propose a method to constrain the initial radius of a baryonic outflow where it is launched  ($R_0$) without the need of identifying an ideal thermal spectrum. We then apply it to GRB 090510 and get that $R_0 \lesssim 6.5\times 10^{6}(\Gamma_{\rm ph}/2000)^{-4}$ cm, suggesting that the central engine is a black hole with a mass $< 22~M_\odot(\Gamma_{\rm ph}/2000)^{-4}$, where $\Gamma_{\rm ph}$ is the bulk Lorentz factor of the outflow at the photospheric radius.
\end{abstract}


\setlength{\parindent}{.25in}

\section{Introduction}
Since the discovery of the afterglow and then the measurement of the redshift in 1997, our understanding of Gamma-ray Bursts (GRBs) has been revolutionized \citep[see][for reviews]{Piran04,Zhang04}. As usual
some aspects are understood better than others. For example the late time ($t>10^{4}$ s) afterglow emission is likely dominated by the radiation of the electrons accelerated by external forward shock while the earlier afterglow emission may consist of at least two components, including that powered by the prolonged activity of the central engine and the external shock emission \citep[e.g.,][]{Fan05a,Meszaros06,Zhang06,Zhang07}. The origin of the prompt $\gamma-$ray emission is less clear. Widely discussed scenarios include the internal shock model, the internal magnetic energy dissipation models, and the photospheric models. The GRBs' central engine is either a stellar black hole surrounded by a hyper-accreting disk \citep{MacFadyen99} or a quickly rotating magnetar \citep{Usov92}. For a magnetar-like central engine the outflow is expected to be Poynting-flux dominated.
In the case of a stellar black hole surrounded by a hyper-accreting disk, the outflow could be either baryonic or Poynting-flux dominated, depending on the energy extraction process \citep[see][for reviews]{Piran04,Zhang04}. Such a kind of central engine is usually characterized by the mass and the spin of the black hole, and the rate of accretion onto the black hole. Our current knowledge of these physical parameters is mainly from the numerical simulations since current electromagnetic data alone can not break the degeneracies between the parameters (see section \ref{sec:3-1}). For some specific long bursts (for which the duration is longer than $\sim 2~{\rm s}$) the situation is better. The modeling of their associated supernovae sheds some light on the mass of the central remnant \citep[e.g.,][]{Deng05}. However, for short GRBs (with a duration less than $2~{\rm s}$) no bright associated-supernova has been detected. In this work we investigate whether it is possible to make some progress with some specific assumptions. We concentrate on the binary neutron star merger model which has been supported by the host galaxy observations and by the non-detection of accompanying bright supernova for some short bursts \citep[e.g.,][]{Gehrels05,Fox05,Barthelmy05,Berger05,Hjorth05}. The other possibility that some short events might have a massive star origin \citep[e.g.,][]{Zhang03,Fan05,Zhang09,Virgili11,Panaitescu11} won't be addressed.

Besides the physical parameters of the central engine, the initial radius of the outflow  where it is launched  ($R_0$) is an important parameter revealing the physical process taking place at the center of the burster. In the collapsar scenario, the interaction of the accelerated/cooled ejecta with the envelope material may give rise to a re-born hot fireball \citep[e.g.,][]{Lazzati09} and then the derived $R_0$ marks the site of the interaction, which is much larger than the initial size of the ejecta. If such an interaction is ignorable, $R_0$ imposes an independent though rough constraint on the mass of the central black hole.
For a baryonic outflow the acceleration and the energy dissipation processes are well understood \citep{Piran93,Meszaros93}. With a reliable thermal component identified in the prompt spectrum, it is possible to constrain the bulk Lorentz factor and $R_0$ of the shells \citep{Peer07}.  Such a goal was achieved for some long GRBs, in particular GRB 090902B, and a typical $R_0 \sim 10^{8}-10^{9}$ cm was inferred \citep[e.g.,][]{Peer07,Ryde10}. For short GRBs, no reliable thermal spectrum component has been identified. A unique candidate is GRB 090510 characterized by a very soft MeV spectrum and a rather hard GeV radiation component \citep{Gao09}. But the peculiar spectrum of this short burst, unlike the long event GRB 090902B, can not be reasonably fitted by a thermal component superposed by a power-law component (B. B. Zhang \& B. Zhang, 2011 private communication). Therefore one purpose of this work is to find a way to estimate $R_0$ without the need of identifying an ideal thermal component.

This work is structured as the following. In section 2 we discuss the difficulty of estimating the physical parameters of the central engine with current observational data, and then show that some progress is achievable in the specific double neutron star merger scenario. The mass of the accretion disk of some short GRBs has been estimated. In section 3 we present a method, in which an ideal thermal signature is not needed, to estimate $R_0$ with the assumption that the outflow is baryonic.  We apply the method to GRB 090510 and then constrain $R_0$. Our results are summarized in section 4 with some discussions.

\section{The mass of the accretion disk of some short GRBs}\label{sec:3}
\subsection{The difficulty of estimating physical parameters of the central engine with current limited  data}\label{sec:3-1}
One nascent stellar black hole surrounded by a hyper-accreting disk, the widely adopted central engine of GRBs, is characterized by some important parameters, including the mass and the spin of the central black hole ($M_{_{\rm BH}}$ and $a$, where $a\equiv J/GM_{_{\rm BH}}^{2}c$, $J$ is the angular momentum of the black hole and $c$ is the speed of light), the mass of accretion disk $M_{\rm disk}$ or alternatively the accretion rate $\dot{M}$.

The accretion disk is so hot that the energy loss may be mainly through neutrino and anti-neutrino radiation. The later neutrino and anti-neutrino annihilation may launch a baryonic fireball \citep{Eichler89}. The annihilation luminosity has been extensively investigated in the literature \citep[e.g.,][]{MacFadyen99,Popham99,Liu07,Zalamea10}. In general the luminosity is a function of $M_{_{\rm BH}}$, $a$, $\dot{M}$, and possibly also the vertical structure of the disk \footnote{The same holds for the models in which it is the magnetic process rather than the neutrino process to extract the energy. For example in the Blandford-Znajek process \citep{BZ77,Lee01}, the luminosity of the electromagnetic outflow can be estimated by $L_{_{\rm BZ}}\approx 2.5\times  10^{49}~{\rm erg}~(a/0.5)^{2}(M_{_{\rm BH}}/2.7M_{_{\rm BH}})^{2}B_{\rm H,15}^{2}$, where $B_{_{\rm H}}\sim 1.1\times 10^{15}~{\rm Gauss}~(\dot{M}/0.01~M_\odot~{\rm s^{-1}})^{1/2}R_{_{\rm H},6}^{-1}$ is the magnetic field strength on the horizon and $R_{_{\rm H}}=(1+\sqrt{1-a^{2}})r_{_{\rm g}}/2$.\label{foot-2}}.  In this work we adopt an empirical relation proposed by  \citet{Zalamea10}, which reads
\begin{eqnarray} L_{\nu\bar{\nu}} &\approx & 10^{52}~{\rm erg~s^{-1}}x_{\rm ms}^{-4.8}({M_{_{\rm BH}}\over 3M_\odot})^{-3/2}\nonumber\\
&&\left\{%
\begin{array}{ll}
    0, & \hbox{for $\dot{M}<M_{\rm ign}$;} \\
    \dot{m}^{9/4}, & \hbox{for $M_{\rm ign}<\dot{M}<\dot{M}_{\rm trap}$;} \\
    \dot{m}_{\rm trap}^{9/4}, & \hbox{for $\dot{M}\geq \dot{M}_{\rm trap}$,} \\
\end{array}%
\right.
\label{eq:main}
\end{eqnarray}
where the accretion rate
  $\dot{m}=\dot{M}/M_\odot~{\rm s^{-1}}$, $x_{\rm ms}=r_{\rm ms}(a)/r_{\rm g}$, $M_{\rm ign}=K_{\rm ign}(\alpha/0.1)^{5/3}$, ${M}_{\rm trap}=K_{\rm trap}(\alpha/0.1)^{1/3}$, and $\alpha$ is the viscosity. The coefficients $K_{\rm ign}$
and $K_{\rm trap}$ are functions of the black hole spin $a$. For $a=0$, $K_{\rm ign}=0.071~M_\odot~s^{-1}$ and $K_{\rm trap}=9.3~M_\odot~s^{-1}$.  For $a=0.95$, $K_{\rm ign}=0.021~M_\odot~s^{-1}$ and $K_{\rm trap}=1.8~M_\odot~s^{-1}$ \citep{Chen07}. The radius of last stable orbit $r_{\rm ms}$ is \citep{Bardeen70ApJ}
\begin{equation}
r_{\rm ms}=r_{\rm g}\{3+Z_{2}\mp [(3-Z_{1})(3+Z_{1}+2Z_{2})]^{1/2}\}/2,
\end{equation}
where \[Z_{1}=1+(1-a^{2})^{1/3}[(1+a)^{1/3}+(1-a)^{1/3}],\] and \[Z_{2}=(3a^{2}+Z_{1}^{2})^{1/2}.\] For $a=0$ we have
$r_{\rm ms}=3r_{\rm g}$, while for $a=1$ we have $r_{\rm ms}=r_{\rm g}/2$ or $r_{\rm ms}=9r_{\rm g}/2$ (retrograde).
The retrograde case is irrelevant to the case of accretion disk and will not be discussed in this work any longer.

With the observational data the energy output of the central engine $\dot{E}_{\rm out}$ and hence $L_{\nu\bar{\nu}}$ can be reasonably inferred, which however is not enough to break the degeneracies among the parameters $a$, $M_{_{\rm BH}}$ and $\dot{m}$. As shown in footnote \ref{foot-2}, the same applies to the magnetic process of energy extraction. That's why {\it it is rather hard to constrain the physical parameters of the central engine with current electromagnetic data} and our knowledge of the central engine is mainly from the numerical simulations.  Fortunately, {\it the specific scenario of binary-neutron-star merger could be an exception, for which a preliminary probe is plausible.}

\subsection{An exception: the double neutron star merger scenario}\label{sec:3-2}
In the specific binary-neutron-star merger model, also the leading one, for short GRBs \citep{Eichler89,Nakar07,Lee07}, the mass of the formed black hole and its spin parameter can be relatively reasonably evaluated.

The mass of the central engine is expected to be close to the total mass of the progenitors since the mass ejection during the merger is expected to be tiny \citep{Rosswog99AA} and the mass range of Neutron Stars is relatively narrow. For the 10 neutron-star-binaries well studied so far, the total-mass of the binaries ranges from $2.57M_\odot$ to $2.83M_\odot$ \citep{Kizi11}. Therefore the mass of the nascent black hole formed in the merger is expected to be close to $M_{_{\rm BH}}\sim 2.7M_\odot$.

The spin parameter of the formed black hole can be semi-quantitatively estimated. Following \citet{Lee01} and for simplicity we assume that the double neutron stars have a similar mass $M_{_{\rm NS}}\sim 1.4~M_\odot$, the orbital angular momentum of the binary system is $J_{\rm binary}=\sqrt{GM_{_{\rm NS}}^{3}r_{\rm t}/2}$, where $r_{\rm t}$ is the tidal radius. The spin parameter of the formed black hole can be expressed by $a\equiv xyJ_{\rm binary}/[(2yM_{\rm NS})^{2}G/c]=(x/4y)\sqrt{r_{\rm t}/r_{\rm g,_{\rm NS}}}$, where $r_{\rm g,_{\rm NS}}=2GM_{_{\rm NS}}/c^{2}$, assuming that a fraction ($xy$) of the orbital angular momentum of the neutron star binary goes into the nascent black hole keeping a fraction $y$ of the total mass. For the binary-neutron-star merger scenario, the tidal radius is $r_{\rm t} \approx 6r_{\rm g,_{\rm NS}}$ \citep{Haense91ApJ} and we have $a=0.61x/y$, about $10\%$ smaller than that suggested by \citet{Lee01}. As already mentioned, $y$ is close to $1$. Let's estimate $x$. The gravitational radiation changes the energy of the binary system at a rate $dE/dt\approx 1.9\times 10^{55}(r_{\rm t}/6r_{\rm g,_{\rm NS}})^{-5}~{\rm erg/s}$ and the merger takes a time $\Delta t \sim 1.4~{\rm ms}~(r_{\rm t}/6r_{\rm g,_{\rm NS}})^{4}$ \citep{Haense91ApJ}. The corresponding change of the angular momentum $\Delta J_{\rm binary}\sim (dE/dt)\Delta t/\sqrt{2GM_{_{\rm NS}}/r_{\rm t}^{3}}$. We then have $x=1-\Delta J_{\rm binary}/J_{\rm binary} \sim 0.9$. Therefore we have $a\sim 0.55$, i.e., the formed black hole rotates rapidly.
As found in \citet{Lee01}, for the double neutron stars having different masses $M_{_{\rm NS-1}}$ and $M_{_{\rm NS-2}}$, roughly one has $a \propto {\cal R}\equiv M_{_{\rm NS-1}}M_{_{\rm NS-2}}^{5/6}/(M_{_{\rm NS-1}}+M_{_{\rm NS-2}})^{11/6}$. For the 10 neutron-star-binaries discussed in
\citet{Kizi11}, one finds out that ${\cal R}$ ranges from $0.275$ to $0.282$, i.e., $a$ is insensitive to the mass ratio of the binaries.
The above semi-quantitative analysis is {\it in agreement with} the numerical simulation, in which people found a typical $a\sim 0.78$, weakly depending on the total mass and the mass ratio of the binary neutron stars \citep{Kiuchi09PRD}.

For $a=0.78$, we have $r_{\rm ms}\approx 1.45r_{\rm g}$, $x_{\rm ms}=1.45$ and (for $M_{\rm ign}<\dot{M}<\dot{M}_{\rm trap}$)
\begin{equation}
L_{\nu\bar{\nu}} \approx  2 \times 10^{51}~{\rm erg~s^{-1}} \dot{m}^{9/4} ({x_{\rm ms}\over 1.45})^{-4.8} ({M_{\rm BH} \over 2.7M_\odot})^{-3/2}.
\end{equation}

A typical $M_{\rm disk}\sim 10^{-3}-10^{-1}~M_\odot$ is found in the numerical simulation of the binary-neutron-star coalescence \citep[see][and the references therein]{Kiuchi09PRD}. The binary-neutron-star merger hypothesis will be supported if our $M_{\rm disk}$ estimated with the observational data is within such a mass range. In the foreseeable future the binary-neutron-star merger model for short GRBs could be directly tested by the Gravitational wave data and the mass of the binaries as well as the formed disk could be inferred \citep{Kiuchi10PRL}. Therefore the validity of our following approach will be directly tested.

\subsection{A simple approach and case studies}\label{sec:3-3}
\textbf{\emph{A simple approach.}} Below we take the simplest approach to estimate $L_{\rm \nu\bar{\nu}}$. The isotropic-equivalent kinetic energy of the outflow powering long-lasting afterglow ($E_{\rm k,iso}$) and the opening angle of the ejecta $\theta_{\rm j}$ can be derived from the modeling of the multi-wavelength afterglow data \citep{Panaitescu06}. However, in a good fraction of short GRBs, such a goal is not achievable due to the lack of prompt observations. Fortunately,
for the X-ray emission above both the typical synchrotron radiation frequency and the cooling frequency, the flux is independent of the poorly constrained number density of the medium $n$ and the X-ray luminosity is a good probe of $E_{\rm k,iso}$ \citep{Kumar00}. Following \citet{Fan06} we take
\begin{eqnarray}
 E_{\rm k,iso} &\sim & 10^{53}~{\rm erg}~{\cal L}_{\rm X,46}^{4/(p+2)}({1+z \over 2}) \epsilon_{\rm B,-2}^{-(p-2)/(p+2)}\nonumber\\
 && \epsilon_{\rm e,-1}^{4(1-p)/(p+2)}(1+Y)^{4/(p+2)},
\label{eq:E_kiso}
\end{eqnarray}
where ${\cal L}_{\rm X}$ is the X-ray afterglow luminosity at $t=10$ hours after the trigger of the burst, $\epsilon_{\rm e}$ and $\epsilon_{\rm B}$ are the fractions of shock energy given to the electrons and magnetic field respectively, $Y$ is the Compton parameter, and $p\sim 2$ is the energy distribution index of the shock-accelerated electrons and is constrained by the X-ray spectrum. The convention $Q_{\rm n}=Q/10^{\rm n}$ has been adopted here and throughout this work except for some specific notations

The jet opening angle is estimated to be \citep{Frail01}
\begin{equation}
\theta_{\rm j} \approx 0.076(t_{\rm j}/1~{\rm day})^{3/8}[(1+z)/2]^{-3/8}E_{\rm k,iso,51}^{-1/8}n_{-2}^{1/8},
\label{eq:theta_j}
\end{equation}
where $t_{\rm j}$ is the jet break time. We then estimate the ``intrinsic" power released by the GRB central engine as
\begin{equation}
\dot{E}_{\rm out}\approx (1+z)(E_{\rm \gamma,iso}+E_{\rm k,iso})\theta_{\rm j}^{2}/(2T_{\rm act}),
\end{equation}
where $T_{\rm act}$ is the duration of the activity of the central engine. In reality, $\dot{E}_{\rm out}$ is just a fraction (${\cal F} \lesssim 0.3$) of the total neutrino-antineutrino annihilation luminosity outside the horizon of the rotating black hole, i.e., $\dot{E}_{\rm out}={\cal F}L_{\nu\bar{\nu}}$ \citep[e.g.,][]{Aloy05}. The real duration of the main activity of the central engine might be shorter than the duration of the prompt emission $T_{\rm 90}$ by a factor ${\cal R} \gtrsim 1$, i.e., $T_{\rm act}=T_{90}/{\cal R}$, because different propagation velocities of the front and rears ends will lead to a radial stretching of the ultra-relativistic ejecta \citep[see section 4.1 of][for more details]{Aloy05}. Hence we have (for $M_{\rm ign}<\dot{M}<\dot{M}_{\rm trap}$)
\begin{eqnarray}
\dot{m} &\approx & 0.53~M_\odot~[{{\cal R}(1+z)(E_{\rm \gamma,iso,51}+E_{\rm k,iso,51})\theta_{\rm j}^{2}\over {\cal F}T_{90}}]^{4/9}\nonumber\\
&& ({x_{\rm ms}\over 1.45})^{2.1}
({M_{\rm BH}\over 2.7M_\odot})^{2/3},
\end{eqnarray}
and
\begin{eqnarray}
M_{\rm disk} &\approx & 0.53~{M_{\odot}}~[{(E_{\rm \gamma,iso,51}+E_{\rm k,iso,51})\theta_{\rm j}^{2}\over {\cal F}}]^{4/9}\nonumber\\
&& [{T_{90}\over (1+z){\cal R}}]^{5/9}({x_{\rm ms}\over 1.45})^{2.1}({M_{\rm BH}\over 2.7M_\odot})^{2/3}.
\label{eq:M_disk}
\end{eqnarray}
Since ${\cal R}\gtrsim 1$ and ${\cal F}<1$, their impacts on estimating $M_{\rm disk}$ are partly canceled. For $a=0.6$, we have $x_{\rm ms} \approx 1.9$, $M_{\rm disk}$ given in eq.(\ref{eq:M_disk}) will be enhanced by a factor of  $\sim 1.8$.\\

\textbf{\emph{Case studies.}}
So far we have 10 short GRBs, as listed in Tab.1, having relatively abundant afterglow data, with which we can estimate $\dot{E}_{\rm out}$ and then $M_{\rm disk}$. For simplicity, instead of discussing all bursts one by one, below we focus on a few special events.

{\it GRB 051221A}, a burst with a duration $z=1.4~{\rm s}$, was at a redshift $z=0.5465$ and is distinguished by a long-lasting X-ray flat segment in the afterglow. Such a flat segment could be due to either the energy injection from the central engine \citep{Soderberg06,Burrows06} or the forward shock emission of the emerging wide-component of the two component jet \citep{Jin07}. The afterglow parameters reported in the literature are different. In the analysis we take the parameters obtained in the two-component jet modeling, in which the long activity of the central engine is not needed. If we take the somewhat more conservative estimate $E_{\rm ej}\approx 3\times 10^{49}$ erg
\citep{Soderberg06,Burrows06}, the mass of the accretion disk will be reduced by a factor of $0.6$.

{\it GRB 090510,} a burst at a redshfit $z=0.903$, is the most energetic short event ever recorded and is also remarkable for its long-lasting GeV emission that is likely powered by the external forward shock \citep{Gao09,DeP10,Corsi10}. The self-consistent interpretation of the GeV/X-ray/optical data is not an easy task and the parameters are found to be somewhat unusual. At $t\sim 1200$ s after the trigger of the burst, the X-ray and the optical afterglow emission change the decline behaviors achromatically \citep{DeP10}. Such changes are most likely due to the jet effect, that is, the edge of the outflow enters our line of sight and the visible emitting region can not be approximated as a spherical surface any longer. The jet opening angle is as small as $\sim 0.006$ \citep{Gao09,Corsi10,He10}, which is about one order of magnitude smaller than that of other bursts (see Table 1) or that found in the numerical simulation \citep{Aloy05}. It is unclear how such a narrow collimation is reached. Nevertheless a similar narrow collimation was identified in the afterglow modeling of the naked-eye burst GRB 080319B \citep{Racusin08}.

{\it GRB 100816A,} as a burst at $z=0.8035$, the local duration is $2.8/(1+z)=1.55$ s, consistent with being a short burst. The result of the Swift-BAT spectral lag analysis is also consistent with being a short hard burst, but the error bars are too large to be definitive
 \citep{GCNRep300-1}. With the X-ray light curve we take a jet break time $t_{\rm j}\gtrsim 2\times 10^{5}$ s. {The most valuable information inferred from the afterglow data of this burst is likely the density profile of the circum-burst medium.} The preliminary white band flux decline is $\sim t^{-1.15}$ for $100~{\rm s}<t<10^{4}~{\rm s}$, steeper than the simultaneous $t^{-1}-$like X-ray decline \citep{GCNRep300-1}. The  spectral index of the X-ray afterglow photons is $\approx 1.03^{+0.12}_{-0.16}$, suggesting that the X-ray emission is above the cooling frequency of the forward shock electrons for $p\sim 2$. If the slow-cooling fireball was expanding into the ISM-like medium and the typical synchrotron radiation frequency is below the observer's band,  the optical afterglow emission should decline with the time as $t^{-0.75}-$like, shallower than the X-ray decline, which is {\it at odds} with the data. For a free-wind medium, the optical decline should be $t^{-1.25}$, steeper than the X-ray decline \citep{Zhang04}. {Therefore the current data favor the free-wind medium model}, in which {the progenitor should be a massive star rather than a pair of compact objects.} If our speculation could be confirmed by the careful analysis of available afterglow data reported in GCNs, the collapsar origin of GRB 100816A with an intrinsic duration $\sim 1.4$ s would be established. In turn such a result would be in support of the hypothesis that collapsar can produce short events \citep{Zhang03} and the progenitors of short GRBs are diverse \citep{Fan05,Zhang09}. If so the calculation made in Table 1 on this kind of bursts is likely {\it invalid}.\\

As shown in Tab.1, for about half of short bursts in the sample, the accretion disk has a mass $\sim 0.01-0.1~M_\odot$, well consistent with that found in the numerical simulations of double neutron star merger \citep[e.g.,][]{Rosswog03,Kiuchi09PRD}. For some other events, such as GRB 051221A and GRB 050724, the inferred $M_{\rm disk}\sim~{\rm several}~0.1~M_\odot$ may be a bit massive to form. This puzzle can be solved in either of the following scenarios. One is that the outflows of these short GRBs were launched via some more efficient magnetic processes rather than the neutrino mechanism, as speculated in the literature \citep[e.g.,][]{Rosswog03,Lee07}. With footnote \ref{foot-2}, for $a\approx 0.78$, $M_{_{\rm BH}}=2.7M_\odot$ and $\dot{M}=(0.1,~1.0)M_\odot~{\rm s^{-1}}$ we have $L_{_{\rm BZ}}\sim (100,~10)L_{\nu\bar{\nu}}$. Consequently the disks with $M_{\rm disk}$ about 10 or more times smaller than that presented in Tab.1 may be enough to power these short events. The outflow launched in this way is Poynting-flux-dominated and the prompt emission due to the magnetic energy dissipation should have a high linear polarization degree. The other is that some short GRBs are from the neutron star-black hole merger for which a massive disk is possible. Another possibility that can not be ruled out is that some short events might have a massive star orign. Since {\it our estimated $M_{\rm disk}$ is close to that found in the numerical simulations}, the compact object merger scenario is likely viable, implying that some (possibly a considerable fraction of) short GRBs may be really driven by the coalescence of double neutron stars and are promising gravitational wave radiation sources.

\section{Estimating the initial radius of the outflow}\label{sec:2}
\subsection{The method}
The following approach is partly motivated by \citet{Peer07}. In their work the photospheric radiation taking place in the radiation-dominated phase and matter-dominated phase have been investigated separately while
we solve the problems jointly. Furthermore we focus on constraining $R_0$ in the {\it absence} of an ideal thermal signature, different from what did in the literature.

For a baryonic outflow, most of the initial thermal energy may have been converted into the kinetic energy of the baryons at the end of the acceleration \citep{Shemi90} but a (quasi-)thermal emission component is likely inevitable \citep{Paczynski90}. The (quasi-)thermal emission is mainly from the photosphere at a radius $R_{\rm ph}$ which satisfies the following relations.

Based on its definition, $R_{\rm ph}$ can be expressed as  \citep[e.g.,][]{Paczynski90,Daigne02,Jin10}
\begin{eqnarray}
R_{\rm ph} & \approx & 4.5\times 10^{11}~{\rm cm}~L_{54}\Gamma_{\rm ph,3}^{-2}\eta_{3}^{-1}\nonumber\\
 &\approx & 3.7\times 10^{11}L_{54}f^{-2}\eta_{3}^{-3},
\label{eq:R_ph1}
\end{eqnarray}
where $f\equiv 3\Gamma_{\rm ph}/4\eta$, $\eta$ is the initial dimensionless entropy, and $\Gamma_{\rm ph}$ is the bulk Lorentz factor of the outflow at the photospheric radius.

The photospheric radius is also related to $R_0$. Following \citet{Piran93} and \citet{Meszaros93}, we introduce a $R_o$, at which the bulk Lorentz factor of the outflow
 is $\Gamma_o\sim $ a few. With the parameter
 \begin{equation}
 {1\over D}={\Gamma_o \over \Gamma_{\rm ph}}+{3\Gamma_o \over 4 \eta_o \Gamma_{\rm ph}}-{3\over 4\eta_o},
 \end{equation}
the acceleration calculation yields \citep{Piran93}
 \begin{equation}
 R_{\rm ph}=R_o{(\Gamma_o/\Gamma_{\rm ph})^{1/2}}D^{3/2},
 \end{equation}
where $\eta_o=e'_o/n'_o m_{\rm p}c^{2}\approx \eta/\Gamma_o$.
We then have ${1\over D}={\Gamma_o / \Gamma_{\rm ph}}+{3\Gamma_o^{2} / 4 \eta \Gamma_{\rm ph}}-{3\Gamma_o/ 4\eta}=3(1-f)\Gamma_o/4f\eta+9\Gamma_o^{2}/16\eta^{2}f$, the first term will be dominant as long as $1-f\geq 3\Gamma_o/4\eta$, which is usually satisfied. So we have
  \begin{equation}
  1/D \approx 3(1-f)\Gamma_o/4f\eta,~~~{D\Gamma_o/\eta}\approx 4f/[3(1-f)],
  \end{equation}
with which we get
\begin{equation}
R_{\rm ph}\approx {4\eta \over 3} R_0 {f \over (1-f)^{3/2}}\approx 1.3\times 10^{10}~{\rm cm}~\eta_{3}R_{0,7}{f \over (1-f)^{3/2}},
\label{eq:R_ph2}
\end{equation}
where the relation $R_0\approx R_o/\Gamma_o$ has been used.

Finally the photospheric radius is constrained by the observational data.
Suppose the (quasi-)thermal emission has a temperature $T_{\rm obs}$ and a flux $F_{\rm bb}$,
 we have
$ 4\pi \Gamma_{\rm ph}^{2}R_{\rm ph}^2 \sigma {T'_{\rm ph}}^{4}=L_{\rm bb}=4\pi D_{\rm L}^2 F_{\rm bb}$,
   which can be simplified as
   \begin{eqnarray}
R_{\rm ph} &\approx & [F_{\rm bb}/\sigma T_{\rm obs}^{4}]^{1/2}(1+z)^{-2}\Gamma_{\rm ph}D_{\rm L}\nonumber\\
  &\approx & 1.3\times 10^{11}~{\rm cm}~F_{\rm bb,-4}^{1/2}[{(1+z)T_{\rm obs}\over 1~{\rm MeV}}]^{-2} f\eta_{3}D_{\rm L,28}.
\label{eq:R_ph3}
\end{eqnarray}
 where $T_{\rm obs}\approx \Gamma_{\rm ph}T'_{\rm ph}/(1+z)$ has been taken into account and and $\sigma$ is the Stefan-Boltzmann constant.

Denoting the co-moving thermal energy density and the number density of the outflow at $R_{\rm ph}$
 as $e'_{\rm ph}$ and $n'_{\rm ph}$ respectively, we have the (quasi-)thermal luminosity
$L_{\rm bb}\approx {e'_{\rm ph} \over 4e'_{\rm ph}/3+n'_{\rm ph} m_{\rm p}c^{2}}L$.
Since
 $e'_{\rm ph}=e'_o/D^{4}$ and $n'_{\rm ph}=n'_o/D^{3}$ \citep{Piran93}, we have
 \begin{equation}
 L_{\rm bb}\approx {e'_o \over 4e'_o/3+D n'_o m_{\rm p} c^{2}}L \approx {L\over 4/3+D \Gamma_o/\eta}\approx {3(1-f)\over 4}L.
 \label{eq:L_bb}
 \end{equation}
 With $Y_{\rm bb} \equiv L_{\rm bb}/L <3/4$, the above relations yield
$f=1-4Y_{\rm bb}/3$ and $\eta=3\Gamma_{\rm ph}/[4(1-4Y_{\rm bb}/3)]$. Combing eq.(\ref{eq:R_ph1}), eq.(\ref{eq:R_ph2}) and  eq.(\ref{eq:R_ph3}) we have
\begin{equation}
R_{0}\approx 1.5\times 10^{8}~{\rm cm}~F_{\rm bb,-4}^{1/2}Y_{\rm bb}^{3/2}[{(1+z)T_{\rm obs}\over 1~{\rm MeV}}]^{-2}D_{\rm L,28},
\end{equation}
\begin{equation}
\Gamma_{\rm ph} \approx 10^{3}~(Y_{\rm bb}^{-1}-4/3)^{1/4}F_{\rm bb,-4}^{1/8}[{(1+z)T_{\rm obs}\over 1~{\rm MeV}}]^{1/2}D_{\rm L,28}^{1/4}.\label{eq:Gamma_ph}
\end{equation}
If the baryon loading is so low that at $R_{\rm ph}$ the outflow is still radiation-dominated, most of the initial energy of the outflow will be lost via the thermal radiation and $Y_{\rm bb}\sim 3/4$, for which {\it $\Gamma_{\rm ph}$ can not be reliably inferred,} reflecting the well-established fact that both the observed temperature and the thermal radiation luminosity are constant until most of the initial energy of the outflow has been transferred into the kinetic energy of the particles \citep{Piran93,Meszaros93}. For $Y_{\rm bb}\ll 1$ (i.e., $R_{\rm ph}$ is far above the coasting radius $\sim \eta R_0$), eq.(\ref{eq:Gamma_ph}) suggests $\Gamma_{\rm ph}\propto Y_{\rm bb}^{-1/4}$, in agreement with \citet{Peer07}.

The above two equations finally give
\begin{equation}
R_0 \approx 1.5\times 10^{8}~{\rm cm}~\Gamma_{\rm ph,3}^{-4}(Y_{\rm bb}^{-1}-4/3)Y_{\rm bb}^{3/2}F_{\rm bb,-4}D_{\rm L,28}^{2}.
\label{eq:Basic-1}
\end{equation}
Therefore, if {\it $\Gamma_{\rm ph}$ and $Y_{\rm bb}$ are obtainable with the observational data}, {\it eq.(\ref{eq:Basic-1}) provides us an independent estimate of $R_0$ without the need of identifying a thermal component in the prompt spectrum.} This is  helpful since the physical processes taking place at $R\leq R_{\rm ph}$ may be able to shape the thermal spectrum  so significantly that the identification of an ideal thermal component is very difficult \citep[e.g.,][]{Beloborodov10}. In fact, for short GRBs, no reliable thermal component has been identified so far.
The disadvantage of our approach is that one can only get the time-averaged constraint.

\subsection{Application to GRB 090510}
GRB 090510 was detected by Fermi $\gamma-$ray telescope and {\it Swift} satellite simultaneously. Though the physical origin of the prompt GeV emission is not clear yet, the long lasting GeV afterglow emission is most likely the synchrotron radiation of the external shock \citep[e.g.,][]{Gao09,Ghirlanda10,Corsi10,DeP10,Kumar10}. Supposing the prompt GeV and MeV emission are from the same region,  the observation of GeV photons sets an upper limit on the optical depth for pair production and then suggests a bulk Lorentz factor of the emitting region $> 1200$ \citep{DeP10,Abdo09}.  To interpret the GeV emission at $t>2$ s as the forward shock emission, a higher initial bulk Lorentz factor of the outflow $\Gamma_{\rm int}\gtrsim 1900$ is required \citep{He10,Ghirlanda10,Corsi10}. Obviously $\Gamma_{\rm ph}$ should be larger than $\Gamma_{\rm int}$ (it is straightforward to show that the outflow shells with higher bulk Lorentz factor contribute more to the thermal emission). In the following estimate we take $\Gamma_{\rm ph}\sim \Gamma_{\rm int} \sim 2000$.
The time-averaged spectrum of GRB 090510 in the time interval $0.5-1.0$ s can be nicely fitted by a Band function plus a power-law component. The Band function component has an isotropic-equivalent energy $E_{\rm Band,iso}\approx 7\times 10^{52}$ erg while the very hard power-law component (with a spectrum $F_\nu \propto \nu^{-0.62}$ in the energy range 10 keV$-$10 GeV) has an isotropic-equivalent energy $E_{\rm PL,iso}\approx 5\times 10^{52}$ erg \citep{Abdo09}. The afterglow modeling gives $E_{\rm k,iso} \gtrsim 5\times 10^{53}$ erg \citep{Gao09,He10}. Clearly, only the Band function component may be relevant to the quasi-thermal radiation of the outflow. Hence we have a (quasi-)thermal radiation efficiency
$Y_{\rm bb}\leq E_{\rm Band,iso}/(E_{\rm k,iso}+E_{\rm PL,iso}+E_{\rm Band,iso})\approx 0.1$ and the thermal radiation flux $L_{\rm bb}\lesssim 6\times 10^{-5}~{\rm erg~s^{-1}~cm^{-2}}$.
Substituting these values into eq.(\ref{eq:Basic-1}), we have
\begin{equation}
R_0 \lesssim 6.5\times 10^{6}~{\rm cm}~\Gamma_{\rm ph,3.3}^{-4}Y_{\rm bb,-1}^{1/2},
\label{eq:Main-1}
\end{equation}
which is about $2$ or more orders of magnitude smaller than that reported in the literature \citep{Peer07,Ryde10}\footnote{For one particular burst GRB 090902B, people found a prominent thermal signature in the prompt spectrum. The modeling of these data suggests a typical $R_0 \sim 10^{9}$ cm \citep{Ryde10,Zhang10}. We speculate that {\it the intrinsic outflow was launched at a much smaller radius and was ``choked" by some material at $\gtrsim 10^{9}$ cm} (i.e., {\it the fireball was re-born,} see also Thompson et al. 2007; Ghisellini et al. 2007; Lazzati et al. 2009) for the following two reasons. One is that the cooling of the disk material is dominated by neutrino radiation process, crucial for launching a baryonic outflow, only inside a radius $\sim 10^{8}$ cm \citep{Narayan01}. The other is that at a radius $\sim 10^{9}$ cm, the annihilation of the neutrinos and anti-neutrinos should be very inefficient since the number density of (anti-)neutrinos drops with radius sharply.}, suggesting that most energy has been deposited in a small cavity surrounding the nascent black hole.
Since the collimation of the outflow in the double neutron star merger scenario should be mainly contributed by the interaction with accretion torus \citep{Aloy05}, a small $R_0$ may be necessary to be consistent with the very small opening angle of the ejecta $\theta_{\rm j}$ found in the afterglow modeling.

The GRB ejecta is mainly launched through the pole region of the rotating black hole. Clearly the outflow should be from a site above the horizon surface. Along the pole that means $R_0> r_{\rm g}\equiv 2GM_{\rm BH}/c^2\approx 3~{\rm km}~(M_{\rm BH}/1~M_\odot)$, regardless of the unknown spin of the black hole, where $r_{\rm g}$ is the Schwarzschild radius and $G$ is the gravitational constant. In reality $R_0$ is larger than $r_{\rm g}$ by a factor of a few, as found in the numerical simulations \citep[e.g.,][]{Popham99,Liu07,Zalamea10}.  With eq.(\ref{eq:Main-1}), for GRB 090510 we then have $M_{\rm BH}<22~M_\odot~~\Gamma_{\rm ph,3.3}^{-4}Y_{\rm bb,-1}^{1/2}$, i.e., it is likely a stellar black hole at the center, in agreement with the compact-object merger model.

For other short bursts, due to the lack of a robust estimate of $\Gamma_{\rm int}$ and then $\Gamma_{\rm ph}$, a reliable constraint on $R_0$ is not possible.

\section{Discussion and Conclusion}
The central engine of Gamma-ray Bursts is widely believed to be one nascent stellar black hole surrounded by a hyper-accreting disk. With the current observational data it is however not easy to pin down the physical parameters, for example, the mass and the spin of the central black hole, the accretion rate (or alternatively the accretion-disk mass), and the initial radius of the outflow where it is luanched ($R_0$). The main reason is that the central engine hides deeply behind the electromagnetic-radiation surface and it is hard to break the degeneracies between the parameters with the very limited observational constraints. For some long bursts, in particular GRB 090902B, people get a reliable estimate of $R_0$ with the identified thermal spectrum component in the prompt spectrum. However for short events the lack of a reliable identification of such a component renders a reasonable estimate difficult. In this work  we discuss whether we can estimate the disk mass in the specific scenario of binary-neutron-star merger (see section \ref{sec:3} for details). We also outline how to constrain $R_0$ without the identification of an ideal thermal spectrum component and then applies it to GRB 090510 (see section \ref{sec:2} for details). Our main conclusions are the following:

\begin{itemize}
\item Our semi-analytical estimate suggests that the nascent black hole formed in binary-neutron-star merger scenario rotates very quickly and the spin parameter is insensitive to the initial mass ratio of the double neutron stars (see section \ref{sec:3-2}), in agreement with the results of the recent numerical simulation \citep[e.g.,][]{Kiuchi09PRD}. Together with the finding that there is no significant mass ejection  in the merger process \citep[e.g.,][]{Rosswog99AA}, the mass of the formed black hole $M_{_{\rm BH}}$ as well as the spin parameter $a$ may be reasonably deduced. Consequently a rough estimate of the accretion-disk mass is possible.

    As found in section \ref{sec:3-3}, for about half of short GRBs in our sample, the disk mass is estimated to be $\sim 0.01-0.1~M_\odot$, in agreement with that found in the numerical simulation of the merger of binary-neutron-star. For some other bursts, such as GRB 051221A and GRB 050724, a massive disk ($\sim {\rm several}~ 0.1~M_\odot$) is needed.
    This puzzle can be solved if the outflows of these short GRBs were launched via the more efficient magnetic processes rather than the neutrino mechanism or alternatively these short GRBs were from the neutron star-black hole merger (for which a massive disk is plausible). Since no significant divergence between the disk mass inferred from the observational data and that obtained in the numerical simulation has been found, we suggest  that the compact object merger scenario for a good fraction of short bursts is viable and these events are promising gravitational wave radiation sources.

\item For GRB 090510, the initial radius of the outflow is estimated to be $\lesssim 6.5\times 10^{6}(\Gamma_{\rm ph}/2000)^{-4}$ cm. Such a small $R_0$ suggests that the neutrino-antineutrino annihilation products were mainly deposited in a small cavity surrounding the nascent black hole, as expected. Moreover, the small $R_0$ imposes a constraint on the mass of the central engine $M_{_{\rm BH}}< 22(\Gamma_{\rm ph}/2000)^{-4}~M_\odot$, consistent with the compact-object merger model.
\end{itemize}

With the future short-burst-associated gravitational wave data, the binary-neutron-star merger model will be directly tested. Moreover the formation process of the disk, total mass and the mass ratio of the double neutron stars involved in the merger and the mass of the formed disk can be well constrained \citep{Kiuchi10PRL,Kobayashi02}. Consequently the validity of our simple approach outlined in section \ref{sec:3} will be unambiguously tested. Considering that the  magnetic process is usually much more efficient than the neutron-antineutrino annihilation to extract the energy, combing the derived $a$, $M_{_{\rm BH}}$ and $M_{\rm disk}$ with $\dot{E}_{\rm out}$, the nature of the outflow-launching-process (magnetic or neutrino-antineutrino annihilation) will be reliably probed, too. For example, if the $M_{\rm disk}$ inferred from the gravitational wave data is so small that can not produce the observed burst via neutrino mechanism, the magnetic outflow-launching-process will be favored.

Finally we'd like to point out that for the possible short event GRB 100816A, the preliminary afterglow data reported in \citet{GCNRep300-1} tentatively favor the free-wind medium model, in which the progenitor should be a massive star rather than a pair of compact objects. Careful analysis of available optical/infrared afterglow data is thus encouraged. If our speculation has been confirmed, the collapsar origin of GRB 100816A with an intrinsic duration $\sim 1.4$ s would be firmly established, in support of the hypothesis that collapsar can produce short events and the progenitors of short GRBs are diverse.

\section*{Acknowledgments}
We are grateful to the anonymous referees for constructive comments, Z. G. Dai and Y. F. Huang for discussion, and S. Kobayashi, W. H. Lei and T. Liu for communication. This work was supported in part by a special grant from Purple Mountain Observatory, the National Natural Science Foundation of China (grants 10973041, 10621303, and 11073057) and Chinese Academy of Sciences and National Basic Research Program of China (grants 2007CB815404 and 2009CB824800).

\clearpage

\begin{table}
\caption{Short GRBs and the mass of the accretion disks.}
\begin{center}
\begin{tabular}{|cl|l|c|c|c|c|c|c|c|c|c|}
\hline
&GRB & $T_{90}(s)$ & z & $E_{\rm \gamma,iso}$ (erg) & $E_{\rm k,iso}$ (erg) & $\theta_{\rm j}$ (rad) &  $M_{\rm disk}~(M_{\odot}$)$^{e}$ & References \\
\hline
&050509B  & 0.05    &  0.2248 & $4.5\times 10^{48}$ & & & 0.02 & 1  \\
&050709  & 0.07    &  0.16 & $6.9\times 10^{49}$ & $3.7\times 10^{50}$ $^a$ & 0.21 &  0.03 & 1--2  \\
&050724   & 3   &  0.257 & $4\times 10^{50}$ & $6.0\times 10^{50}$ $^a$ & 0.2 & 0.37 & 1--2  \\
&051221A  & 1.40    &  0.5465  & $2.4\times 10^{51}$ & $1.0 \times 10^{52}$ $^{b}$ & 0.10 & 0.46 & 3--4  \\
&061006   & 0.4 &  0.4377  & $2.1\times 10^{51}$ & $1.8 \times 10^{51}$ $^{c}$ & $\sim 0.11$ & 0.10 & 5--6 \\
&070714B  & 3   &  0.9224 &  $1.2\times 10^{51}$ & $3.6\times 10^{51}$ $^{c}$ & $\geq 0.08$ $^{d}$ & $\geq 0.24$ & 7\\
&071227   & 1.8 &  0.381  &  $5.8\times 10^{50}$ & $5\times 10^{50}$ $^{c}$ & $\geq 0.09$ $^{d}$ & $\geq0.13$ & 8  \\
&090426   & 1.25    &  2.609    & $3\times 10^{51}$ & $8.7\times 10^{52}$ & $0.06$ $^{d}$ & $0.30$ & 9\\
&090510   & 0.30    &  0.903 &  $1.2\times 10^{53}$ & $5\times 10^{53}$ & $0.006$  & 0.06 & 10--12 \\
&100816A   & 2.8    &  0.8035 &  $5.8\times 10^{51}$ & $1.1\times 10^{52}$ $^{c}$ & $\geq 0.01$ $^{d}$  & $\geq 0.07$ & 13 \\

\hline
\end{tabular}
\end{center}
\begin{minipage}{18cm}
$^a$ We take the lower value obtained in \citet{Panaitescu06}. \\
$^b$ There was a flat segment in the X-ray afterglow and its origin is still unclear. Here we take an $E_{\rm k,iso}\sim 10^{52}$ erg required in the two-component jet model \citep{Jin07}. \\
$^c$ This parameter is estimated by eq.(\ref{eq:E_kiso}) by taking $\epsilon_{\rm e} \sim 0.1$, $\epsilon_{\rm B}\sim 0.01$ and $Y\sim {\cal O}(1)$. \\
$^{d}$ The jet opening angle is estimated by eq.(\ref{eq:theta_j}). For GRB 090426, a jet break time $t=0.4$ day (Nicuesa Guelbenzu et al. 2011) and $n\sim 10~{\rm cm}^{-3}$ (Xin et al. 2011) have been adopted. For other events, we take the time of the last {\it Swift} XRT detection, reported at http://www.swift.ac.uk/xrt$_{-}$curves/, as $t_{\rm j}$ and $n\sim 0.01~{\rm cm^{-3}}$ to set a lower limit on the half opening angle. \\
$^{e}$ We estimate $M_{\rm disk}$ with eq.(\ref{eq:M_disk}), adopting $M_{\rm BH}=2.7~M_\odot$, $a=0.78$, ${\cal F}=0.3$ and ${\cal R}=1$.\\
---References:
(1)~\citealt{Fox05}; (2)~\citealt{Panaitescu06}; (3)~\citealt{Soderberg06}; (4)~\citealt{Jin07}; (5)~\citealt{Berger07}; (6)~\citealt{GCN5710}; (7)~\citealt{Cenko08}; (8)~\citealt{Caito10}; (9)~\citealt{Xin10};  (10)~\citealt{Abdo09}; (11)~\citealt{He10}; (12)~\citealt{Gao09}; (13)~\citealt{GCNRep300-1}.
\end{minipage}
\label{tab:sample}
\end{table}

\end{document}